**Charge-state lifetimes of single molecules on ultrathin insulating films**

Katharina Kaiser*[1] #, Leonard-Alexander Lieske[1], Jascha Repp[2], Leo Gross*[1]

[1] IBM Research Europe – Zurich, Säumerstrasse 4, 8803 Rüschlikon, Switzerland

[2] Department of Physics, University of Regensburg, Universitätsstraße 31, 93053 Regensburg, Germany

# Present address: Université de Strasbourg, CNRS, IPCMS, UMR 7504, F-67000 Strasbourg, France

*Corresponding authors. E-mail: katharina.kaiser@ipcms.unistra.fr, LGR@zurich.ibm.com

In scanning tunneling microscopy (STM) experiments of molecules on insulating films, tunneling through molecular resonances implies transiently charging the molecule. The transition back to the charge ground state by tunneling through the insulating film is crucial, for example, for understanding STM-induced electroluminescence. Here, using STM, we report on the charge-state lifetimes of individual molecules adsorbed on NaCl films of different thicknesses on Cu(111) and Au(111). To that end, we approached the tip to the molecule at resonant tunnel conditions up to a regime where charge transport was limited by tunneling through the NaCl film. The resulting saturation of tunnel current is a direct measure of the molecule's charge-state lifetime, thus providing a means to study charge dynamics and, thereby, exciton dynamics. Comparison of anion and cation lifetimes on different substrates reveals the critical role of the level alignment with the insulator's conduction and valence band, and the metal-insulator interface state.

Introduction

Single-molecule charge transfer plays a significant role in many areas, from molecular electronics [1,2], single-molecule light emission [3–9], photocurrent generation [10] to natural processes such as photosynthesis [11]. Over the past 20 years, the tremendous progress in high-resolution scanning probe microscopy has facilitated the investigation of single molecules with charge state control [5,8,12–17]. In scanning tunneling microscopy (STM), this was enabled by introducing a thin insulating film as a decoupling layer between molecule and metallic substrate, preventing hybridization between molecule and substrate while still allowing charge transfer and pre-serving a sufficient conductance for STM [3,18–20]. This facilitates, for example, mapping molecular ion resonances, which is based on a transient change in the charge state of the molecule [19]. For the investigation of molecular electroluminescence in STM-induced luminescence (STML) experiments, the decoupling layer serves two purposes: It reduces luminescence quenching from the metallic substrate [3,21–23], and, due to the finite lifetime of charged species, it enables an exciton formation mechanism based on subsequent charge transfer from tip and sample [7,24].

Such experiments on thin insulating films have in common that at sufficiently high bias voltages, sequential tunneling through a molecular resonance sets in. In this two-step sequential tunneling process, the molecule is transiently charged by a tunneling event between molecule and tip, followed by a tunneling event between molecule and metallic substrate. In almost all cases, the former tunneling event involving the tip is the current-limiting factor, such that little is known about the rate of the second tunneling transition involving the substrate. However, the latter can be critical for the interpretation of experimental results. For example, the aforementioned sequential tunneling process can – depending on the level alignment – lead to the formation of an excited state [3,7,8], which can subsequently decay under the emission of a photon. The excitation mechanism is fundamentally different from optical excitation because it entails a two-step process [8,9,25], *i.e.*, charging from the tip and subsequent charge transfer to the substrate. Hence, the entire cycle of electroluminescence including the emission of a photon already involves (at least) three transitions with their respective rates. In addition, the creation of the exciton by charge transfer competes with the neutralization of the molecule to its neutral ground state involving even a fourth rate. Thus, the charge-state lifetime of the adsorbed molecule needs to be taken into account in any consideration of dynamics in STML experiments, all-electric pump-probe measurements of excited states [24,26–30] as well as yields in photocurrent generation [10].

The average elapsed time between charging by tunneling between tip and molecule, and neutralization by charge transfer between molecule and substrate depends on the tunneling probability between molecule and metallic substrate and thus on the thickness of the insulating film [15,16,31,32]. Although this time is a property of the entire system and occurs in an out-of-equilibrium situation, in the following we refer to this quantity as the charge-state lifetime.

One way of experimentally determining charge-state lifetimes has been demonstrated for Cl-vacancies in NaCl films of various thicknesses [31]. Analogously to surface-adsorbed molecules, these defects exhibit electronic resonances and can be transiently charged at sufficiently high bias voltages. At resonance, the system represents a double-barrier tunnel junction with one barrier corresponding to tunneling between tip and defect (vacuum barrier) and the other corresponding to tunneling between defect and metal substrate (NaCl barrier). At tunnel conditions in typical STM experiments, the charging step by tunneling through vacuum is the rate-limiting step and therefore determines the measured current $I$. In this (usual) regime, the current $I(z)$ increases exponentially with decreasing tip height $z$. At close distances, however, the time for charging the neutral defect by tunneling through vacuum ($\tau_c$) can become smaller than the time to discharge the defect through the insulating film (*i.e.*, the charge-state lifetime $\tau_d$), and thus, $I(z)$ reaches a $z$-independent saturation current $I_{sat}$ for small $z$. Specifically, at the first (positive and negative) ion resonance, the doubly-charged state is energetically not available due to Coulomb repulsion, such that in this regime, $I$ is limited by the defect's charge-state lifetime $\tau_d$ and for small $z$ no longer depends on $z$ [22,29–32]. Hence, under these conditions, one can directly deduce $\tau_d = q/I_{sat}$, with the elementary charge $q$, from the saturation current $I_{sat}$. Since Cl-vacancy states are localized within the top-most NaCl layer, possess an $s$-wave character as well as strong lateral confinement, and have no occupied state in the relevant energy range, it is not straightforward to generalize the results from vacancies to molecules.

Here, by extending this approach to single molecules, as sketched in Fig. 1, we investigated charge-state lifetimes of ZnPc and $H_2$Pc molecules, both of which are frequently used model systems in STML experiments [4,29,33], adsorbed on 3-5 monolayers (ML) of NaCl on Au(111), *i.e.*, NaCl/Au(111) and Cu, *i.e.*, NaCl/Cu(111). Comparing results on the different work function surfaces Au(111) and Cu(111) allows us to shift the molecular electronic states with respect to the sample's electrochemical potential in order to probe the effect of different possible tunnel channels for the neutralization event.

Results

We performed current-versus-distance spectroscopy $I(z)$ within the electronic resonances of ZnPc and $H_2$Pc molecules adsorbed on 3 to 5 monolayer (ML) thick NaCl films on Cu(111) and Au(111) substrates. To that end we applied voltages $V$ at and, in absolute values, up to few 100 mV above the respective peak positions in d$I$/d$V$, $V_{PIR}$ and $V_{NIR}$ (see Methods as well as Fig. 2e, f and Fig. S2 in the Supporting Information for scanning tunneling spectra and additional information). Figures 2a and b show $I(z)$-curves recorded above a lobe of a ZnPc at the first negative ion resonance (NIR) and positive ion resonance (PIR), respectively, for different NaCl film thicknesses $d_{NaCl}$.

As in the case of Cl-vacancies, the spectra in Fig. 2a, b show two distinctively different regimes: At large $z$, the $I(z)$ spectra exhibit an exponential $z$-dependence; the current-limiting tunnel process is the tunneling through vacuum between tip and molecule. For small $z$, $I(z)$ shows a plateau at the saturation current $I_{sat}$. This shows that at the electronic resonances tunneling between tip and metal substrate is governed by a two-step tunnel process *via* the molecule, whereas the contribution of direct tunneling, which should increase exponentially with decreasing $z$, is negligible. In addition, in the regime of saturation the tunnel current is governed by tunneling through the NaCl barrier, and the molecule is charged most of the time.

The lateral dependence of the current can be used to experimentally verify if the saturation observed in $I(z)$ is due to the effect described above. If the current was indeed limited by molecule-to-substrate tunneling, it should be also largely independent of the lateral tip position. Only if the tip was placed laterally off of the entire molecule, the current should drop. Constant-height (c.h.) STM images recorded at resonance, shown in Fig. 2c for ZnPc adsorbed on 5 ML NaCl recorded at the PIR, *i.e.*, $V$ = -1.2 V, show a transition to this regime. For a larger tip-height offset of $\Delta z$ = 6.5 Å, the c.h. STM map shows a maximum current of about 0.5 pA, which is smaller than the saturation current $I_{sat}$, and a significant variation of the tunnel current is observed as a function of the lateral tip position above the molecule, as usual [19].

Upon decreasing $z$, the current increases in all regions above the molecule, however, only until reaching the saturation current $I_{sat}$ for ZnPc/NaCl(5 ML)/Au(111) of about 0.9 pA, compare with Fig. 2b. At $\Delta z$ = 4.5 Å, we obtain a "flat-top" contrast with the saturation current reached at all positions above the molecule, and nodal planes can no longer be observed. This confirms that $\tau_d$ is independent of the position at which the charge had been attached to the molecule. Such current behavior in c.h. STM maps is reproduced for different film thicknesses and resonances, see Supporting Information Fig. S1. Bias voltage dependent measurements reveal that $\tau_d$ remains the same within a few 100 mV around the peak voltage of the resonance and several 100 mV above it (see Fig. S3a, b). By changing the total applied bias voltage (and tip height), the voltage drop across NaCl also changes by a small amount [34]. $I_{sat}$ being independent of $V$ (for voltages associated to a certain ion resonance, see Fig. S3a, b) and of the tip height $z$ (Fig. 2a, b) indicates that the charge-state lifetime does not vary significantly as a function of the voltage drop across NaCl for a specific ion resonance. A more detailed discussion of the voltage dependence and the partial voltage drops in vacuum and NaCl is given in the Supporting Information S2.

The $I(z)$-curves shown in Fig. 2a, b can be fitted assuming a total rate $\Gamma_{tot}$ for the entire charge transfer process between tip and metal substrate that comprises the charging time $\tau_c$ of the neutral molecule by tunneling through the vacuum barrier, with an exponential $z$-dependence, and the charge-state lifetime $\tau_d$ by tunneling through the NaCl film that is independent of $z$.

$$\frac{1}{\Gamma_{tot}}(z) = \tau_c(z) + \tau_d = \frac{1}{\Gamma_{vac}(z)} + \tau_d = \frac{1}{\Gamma_{vac}^0 \exp(-2\kappa_{vac} \cdot z)} + \tau_d$$

$$I(z) = q \cdot \Gamma_{tot} = \frac{\Gamma_{vac}^0 \cdot q}{\exp(2\kappa_{vac} \cdot z) + \Gamma_{vac}^0 \cdot \tau_d}$$

Here, $\Gamma_{vac}(z)$ is the rate at which the neutral molecule is charged via the tip through the vacuum barrier (as a function of $z$) and $\Gamma_{vac}^0$ is that rate for $z = 0$. $\Gamma_{vac}^0$ as well as $\kappa_{vac} = \frac{\sqrt{2m_e \phi_{vac}}}{\hbar}$ are fitting parameters, with the electron mass $m_e$ and the effective vacuum barrier height (between molecule and tip) $\Phi_{vac}$. The fits to the experimental $I(z)$-curves are shown as dashed red lines in Fig. 2a and b, respectively.

Figure 3 shows the extracted saturation currents and related charge-state lifetimes $\tau_d$ of anions and cations of ZnPc and H$_2$Pc on NaCl films on Au(111) and Cu(111) as a function of $d_{NaCl}$. The charge-state lifetimes $\tau_d$ range from around 50 ps (3ML NaCl/Cu(111)) to 20 ns (5ML NaCl/Au(111)). The previously reported charge-state lifetimes of the anionic charge states of Cl-vacancies in NaCl [31] are about one order of magnitude smaller (on the order of 1 ns for 5ML NaCl) compared to the charge-state lifetimes of the investigated negatively charged molecules on 5 ML NaCl. Aside from other effects, *e.g.* different barrier heights and energetic positions of the resonances, this relates to the probed Cl-vacancies being located within the top monolayer of the NaCl-film, whereas molecules are adsorbed on top of the NaCl film with an adsorption height of approximately 0.3 nm [35].

As conventional $I(z)$ spectroscopy can be used to extract $\Phi_{vac}$ [36,37], the slope $m = \ln(I_{sat})/d_{NaCl}$ provides access to the effective NaCl barrier height $\Phi_{NaCl}$. Comparing the data in Fig. 3, we make two important observations: First, the slopes $m$ for both molecules and at both ion resonances (PIR and NIR) are very similar, with $m \approx -2.9$ ML$^{-1}$ for Cu(111), see Fig. 3a, and $m \approx -3.4$ ML$^{-1}$ for Au(111), see Fig. 3b. That is, for both cations and anions, $I_{sat}$ decreases ($\tau_d$ increases) by a factor of 18 for Cu(111) and by a factor of 30 for Au(111), per added monolayer NaCl. The corresponding extracted effective NaCl barrier-heights are $\Phi_{NaCl/Cu} \approx 0.9$ eV on Cu(111) and $\Phi_{NaCl/Au} \approx 1.3$ eV on Au(111). The full list of slopes and effective barrier heights is reported in the supporting information in Table S1. Second, while on Cu(111), the lifetimes of cations (PIR) and anions (NIR) are similar (see Fig. 3a), on Au(111), the lifetime of the cations is about one order of magnitude longer compared to that of the anions (NIR) (comparing same NaCl layer thicknesses, see Fig. 3b).

The first observation, namely the similar $m$ for positive and negative ion resonance, indicates that the NaCl effective barrier heights $\Phi_{NaCl}$ for the transitions from anion and cation to the neutral state are similar. This is different compared to tunneling through vacuum, where the barrier height increases with increasing energy difference to the vacuum level and therefore is smaller for the NIR compared to the PIR.

In the following, we will discuss these observations based on the energetic alignment of potential tunnel channels that may be involved in the neutralization process with respect to the substrate states (Fig. 4). Note that the picture presented in Fig. 4 is a sketch with estimations based on previous experiments (see Fig. S5), which could be refined and better quantified by future experiments and theoretical investigations.

In general, neutralization of a charged species in a $D_0^{+/-}$ state can, depending on the level alignment between molecule and metal substrate, proceed *via* several channels, *i.e.*, in addition to transitions to the neutral ground state ($S_0$) transitions to excited states (*e.g.*, $T_1$, $S_1$) could be energetically allowed [3,4,9,38]. Because only a small fraction of the applied bias drops in the NaCl film, about 10-20% for the used geometries (see, *e.g.*, [15] and Supporting Information S2), by changing the applied bias voltage we can shift the molecular levels only within a very limited energy window with respect to the substrate's electrochemical potential. Instead, we can use metal substrates with different work functions to compare situations where only the channel to the neutral ground state is energetically allowed, to situations where also channels to excited states are possible. Although the exact positions of the anion's/cation's electronic levels are not accessible in STM on ultrathin insulating films, we can deduce for which cases excited states can be formed in the neutralization step, *e.g.*, from comparison to STML measurements and comparing the energies of the PIR and NIR with excited state energies (*i.e.*, energies of optical transitions). This information allow us to draw an estimated picture of the level alignments between the transiently charged molecules and the substrate. We assume that on Cu(111) the anion, formed at the NIR at about $V_{NIR} = 1$ V, only neutralizes to the ground state $S_0$, whereas the cation, formed at the PIR at about $V_{PIR} = -$ 2.2 V, can also form the excited $S_1$ and $T_1$ state upon neutralization. On the other side, we assume that on Au(111) the cation, formed at the PIR at about $V_{PIR}$ = -1.1 V, only neutralizes to $S_0$, whereas the anion, formed at the NIR at about $V_{NIR}$ = 2.1 V, can also form the excited $S_1$ and $T_1$ state [5,38]. For a more detailed discussion see the Supporting Information S3 and Fig. S5. This energetic alignment of the different channels is considered and reflected in the schematics shown in Fig. 4, and it allows us to discuss how the lifetime changes when different channels are open for neutralization. Note that the level alignments in Fig. 4 depict the molecular resonances after the charging event *via* the tip, which differ from those of the neutral molecule (Fig. 2e, f) due to, *e.g.*, Coulomb repulsion, lifted spin degeneracy and reorganization energy [7,15].

The experimentally observed similar *m* for anion and cation indicates similar effective barrier heights for the charged-to-neutral transitions. If here the effective barrier height simply corresponded to the respective energetic difference to the vacuum level (or another single fixed energy level, *e.g.*, the conduction band minimum or valence band maximum), the extracted effective barrier height $\Phi_{NaCl}$ would be significantly different for the PIR (neutralization of the cation) compared to the NIR (neutralization of the anion). Note that in general, for the NIR (PIR) neutralization occurs by electrons tunnelling at energies above (below) the chemical potential of the metal sample, *i.e.*, to unoccupied (from occupied) sample states. We hypothesize that the similar *m* result from the effective NaCl barrier height $\Phi_{NaCl}$ being determined by the relative energetic position of the tunnel channels with respect to the band structure of the NaCl film. It has been shown that for tunnel processes involving states that are energetically near the VBM, hole tunneling, for which the tunnel barrier height is given with respect to the VBM and thus decreases with decreasing carrier energy, can dominate [39]. Especially for the cation (Fig. 4b, d), the energetic difference of one of the tunnel channels to the VBM becomes comparable to the energetic difference of tunnel channels to the CBM and vacuum level in other cases, which could explain its similarly small *m* compared to the anion (Fig. 4a, c). However, since the different neutralization processes involve tunnel channels with different energetic alignment with respect to the substrate's states and thus with respect to CBM and VBM (see Fig. 4), it seems unlikely that the effective barrier heights result from the energetic separation of these channels to either VBM or CBM alone. Instead, the similar *m* suggest that the effective barrier height for the neutralization of the ionic molecules is nearly energy independent for the different processes studied in this work. In fact, tunnelling through a solid is correctly described by a complex band structure [40–42]: inside the band gap the wave vectors become imaginary and the wave functions decay exponentially into the bulk. Such imaginary wave vectors adequately describe the tunnelling and thereby translate into an effective tunnelling barrier. Complex band structure calculations for the wide-band-gap insulator MgO suggest that deep inside the band gap the imaginary wave vector becomes nearly independent of energy [40–42]. We expect the complex band structure of NaCl to be qualitatively similar, which would explain our observation of *m* being nearly independent of energy. Note that, in direct vicinity to CBM or VBM, we expect the effective barrier height to be smaller and exhibit a stronger energy dependence compared to deep in the band gap, as observed for doubly charged Cl vacancies [31]. The potential landscape being influenced by the different transient charge states of the molecule as the initial state of the tunnelling process through NaCl may further influence the effective barrier heights.

Since we find similar slopes *m* in ln($I_{sat}$) for the decharging of anion and cation from the sample, the second observation, *i.e.*, the cation lifetimes on Au(111) being about one order of magnitude longer than that of the anions, cannot be explained by differences in the effective barrier height for the involved tunneling events. Instead, the observed differences can be rationalized by considering the sample's local density of states (LDOS) at the energy of the ion resonances. The interface state (IS) band that descends from the Shockley surface state of noble metal (111) surfaces [43,44] upon adsorption of a dielectric [45–48] extends the sample's LDOS, and thus reduces the length of the tunnel path through NaCl for transitions including the IS, and modulates it [34,46,49,50]. The onset of the NaCl/metal IS for Cu(111) is at about *V* = -220 mV and for Au(111) at *V* = -270 mV [45,46]. Figure 4 shows schematically in a single-electron picture how the IS potentially contributes to tunneling through the NaCl barrier: It contributes for all systems studied, except for the cation on Au(111), where presumably only tunneling from the substrate at energies below the onset of the IS can neutralize the cation, see Fig. 4d. We propose that this is the main reason for the comparably long lifetimes of the cations on Au(111). In contrast, for the cations on Cu(111), tunneling from the substrate to neutral excited states is energetically possible [5,7,8,38], opening channels at IS energies, see Fig. 4b. In addition, the number of accessible tunnel channels might influence the lifetime. Differences in the spatial extension of the singly occupied orbitals' wavefunctions will also affect the relative contributions of the different channels. The latter argument is not considered in Fig. 4.

In summary, on both surfaces, the neutralization of the anion and cation by tunnelling through the NaCl film exhibits very similar effective barrier heights, although the energetic separation of the involved tunnel channels with respect to the vacuum level, the VBM, and the CBM is very different. This leads us to the conclusion that $\Phi_{NaCl}$ results from the relative energetic position of the tunnel channels inside the band gap of the NaCl film. Deep inside the band gap, $\Phi_{NaCl}$ does not seem to vary significantly with the relative energetic position of the tunnel channels. In addition, on Au(111), the longer lifetimes observed at the PIR compared to the NIR – despite similar effective barrier heights – indicate tunnelling across different distances. This can be rationalized by the increased LDOS due to the IS, that contributes to at least one channel in all situations (Fig. 4a, b, c) except for the PIR on Au(111) (Fig. 4d).

As a side remark, if several channels are open, *i.e.*, as for the PIR on Cu(111), see Fig. 4b, and for the NIR on Au(111), see Fig. 4c, the fastest channels should dominantly govern the charge-state lifetime and the measured effective barrier height. However, in our experiment we cannot separately measure the rates of competing channels. Based on our results and arguments discussed above we would expect that the channels that involve the IS, sketched bold in Fig. 4, are faster and thus dominant. STML showing light emission from the $S_1$ state for the NIR on Au(111) [5,17,38], indicates that out of the three channels that involve the IS, shown in Fig. 4c, the channel that is lowest in energy contributes at least significantly.

We note that momentum conservation as well as the wave vector parallel to the surface ($\vec{k}_{||}$) can additionally influence the tunneling probabilities [51–54], but estimating this influence based on the momenta of the involved states [46,55,56] is beyond the scope of this work.

At voltages sufficiently exceeding the first electronic resonances, additional tunnel channels can be accessed [31,57,58]. Figure 5a shows $I(z)$ spectroscopy of ZnPc on NaCl(5 ML)/Au(111) at different negative sample voltages. Very similar $I(z)$ spectra and saturation currents $I_{sat}$ are measured for $V$ from -1.6 V to -2.4 V. At $V$ = -2.5 V, however, the current shows a much less pronounced plateau at about $z$ = -0.5 Å and then further increases with decreasing $z$. For $z$ < -1.5 Å, the current significantly exceeds the saturation current of $I_{sat} \approx 0.9$ pA measured at less negative voltages. This is also visualized in c.h. STM images at $V$ = -2.5 V at different tip-sample distances (Fig. 5b). For $\Delta z$ > 5.0 Å, the same behavior as for smaller negative voltages is observed, *i.e.*, with decreasing $z$, the current reaches saturation at an increasing number of lateral positions above the molecule until a flat-top current image is observed (at $\Delta z$ = 5.0 Å). Going closer with the tip ($\Delta z$ < 4 Å), we observe that the current increases further at some lateral positions but remains "flat" in other regions above the molecule. Interestingly, the spatial distribution of regions of increased current shows a ring of 12 maxima, as on a clock, with bright maxima at positions 3, 6, 9, 12 o'clock and two equally spaced fainter maxima in between. This corresponds to the shape and symmetry observed for the LUMO rather than the HOMO density, which exhibits three (not two) equally spaced fainter lobes between the bright lobes (see Fig. 5c for comparison).

The contrast that we observe at $V$ = -2.5 V for small $z$ (at $\Delta z$ = 3 Å) could be explained by transitions involving higher-lying states of ZnPc, such as for example higher charge states [57–59], *i.e.*, the dication ($S_0^{2+}$), or trionic states ($D_n^+$) [5,17]. Some potentially accessible states and corresponding transitions are shown in a many-body energy diagram in Fig. S3 in the SI.

The dication, *i.e.*, the doubly positively charged ground state ($S_0^{2+}$), could be accessed by applying sufficiently large currents and bias voltages in a two-electron process, which becomes significant in a regime where the $D_0^+$ → $S_0$ transition (neutralization by tunneling through NaCl) is slower than $D_0^+$ → $S_0^{2+}$ (tunneling another electron to the tip), *i.e.*, at small $z$. We observe a corresponding behavior in the $I(z)$-curves in Fig. 5a and the c.h. STM maps in Fig. 5b, where for $\Delta z$ = 5 Å, we observe the saturation of the tunnel channel corresponding to the $S_0$ → $D_0^+$ transition as well as a subsequent increase of the current at $\Delta z$ < 5 Å.

Due to a non-zero population of $S_0^{2+}$, transitions into other higher-lying states, such as the cation's excited state $D_1^+$, become accessible. Thus, different tunnel channels can contribute to the overall tunnel current [60,61], and the contrast in STM results from a superposition of these channels [60,62]. In the transition $D_1^+$ → $S_0^{2+}$, for example, an electron is removed from the former LUMO (which is singly occupied in $D_1^+$) by tunneling through the vacuum barrier, which could explain the observed contrast in Fig. 5b at $\Delta z$ = 3 Å.

The results in Fig. 5 demonstrate that additional transitions can contribute to the overall tunnel current when increasing the bias voltage, and their relative contribution can be tuned with tip-sample distance. These transitions can play an important role, for example, in the formation of excited states in STML experiments, where exciton formation *via* the singly charged molecule is not always energetically possible and is thus sometimes only observed at higher-lying ion resonances [5,27,63–65].

In conclusion, we reported lifetimes of transiently charged molecules on thin NaCl films on Au(111) and Cu(111) surfaces. Previously, Hanbury Brown-Twiss interferometry and phase fluorometry in combination with STML have been used to access exciton dynamics in single molecules adsorbed on ultrathin insulating films [27,29].

The extracted lifetimes comprise time constants for both excitation and decay of molecular excitons and are of similar magnitude as the charge-state lifetimes reported here (on the order of 500 ps for 3ML NaCl). This indicates that, for certain geometries, the neutralization process from the substrate is the rate limiting process in the formation and decay of molecular excitons and, thus, the previously extracted time constants are likely dominated by the charge-state lifetime. Our results further indicate that the effective NaCl barrier heights for neutralization are governed by the energetic alignment of the tunnel channels inside the band gap of the insulating decoupling layer. Moreover, the energetic alignment of the tunnel channels for neutralization with the interface state significantly impacts the charge-state lifetime. Our results provide an improved understanding of the tunnel processes in these relevant double-barrier tunnel junctions and a quantification of the lifetime of transiently charged molecules, important for understanding excited-state formation by charge attachment in the growing field of STML experiments.

Methods

We performed the experiments in a home-built low-temperature ($T \approx 5$ K) combined STM/AFM system operated under UHV conditions and at a base pressure of $1\times10^{-10}$ mbar. The voltage $V$ was applied to the sample. The metal substrates were cleaned by repeated Ne$^+$ ion sputtering and annealing cycles. NaCl was deposited at sample temperatures between 250 K and 300 K [45]. ZnPc and H$_2$Pc were sublimed onto the cold (T $\approx$ 10 K) substrate from a Si-wafer.

For constant-height STM and $I(z)$ spectroscopy, we approach the tip by the tip-height offset $\Delta z$ from a given STM-controlled setpoint, indicated in each caption. In $I(z)$ spectroscopy the offset for the tip height $z$ is chosen such that at $I = 0.5$ pA, $z$ is 0 Å in every spectrum. Increases in $z$ and $\Delta z$ correspond to increases in tip-sample distance. We define the voltages corresponding to the peaks in d$I$/d$V$ as $V_{PIR}$ and $V_{NIR}$, respectively. The voltages $V$ used in the experiment for probing the lifetime at PIR and NIR were chosen such that, by varying $z$ and thus changing the lever arm for the voltage drop across NaCl, the molecular electronic resonances stay accessible. To this end, we typically used for all tip heights slightly larger voltages (up to few 100 meV) compared to the respective peak positions in d$I$/d$V$ for the larger tip heights probed.

References

[1]  M. A. Ratner, *Introducing Molecular Electronics*, Mater. Today **5**, 20 (2002).
[2]  N. Xin, J. Guan, C. Zhou, X. Chen, C. Gu, Y. Li, M. A. Ratner, A. Nitzan, J. F. Stoddart, and X. Guo, *Concepts in the Design and Engineering of Single-Molecule Electronic Devices*, Nat. Rev. Phys. **1**, 211 (2019).
[3]  X. H. Qiu, G. V. Nazin, and W. Ho, *Vibrationally Resolved Fluorescence Excited with Submolecular Precision*, Science **299**, 542 (2003).
[4]  B. Doppagne, M. C. Chong, E. Lorchat, S. Berciaud, M. Romeo, H. Bulou, A. Boeglin, F. Scheurer, and G. Schull, *Vibronic Spectroscopy with Submolecular Resolution from STM-Induced Electroluminescence*, Phys. Rev. Lett. **118**, 127401 (2017).
[5]  B. Doppagne, M. C. Chong, H. Bulou, A. Boeglin, F. Scheurer, and G. Schull, *Electrofluorochromism at the Single-Molecule Level*, Science **361**, 251 (2018).
[6]  G. Chen, Y. Luo, H. Gao, J. Jiang, Y. Yu, L. Zhang, Y. Zhang, X. Li, Z. Zhang, and Z. Dong, *Spin-Triplet-Mediated Up-Conversion and Crossover Behavior in Single-Molecule Electroluminescence*, Phys. Rev. Lett. **122**, 177401 (2019).
[7]  K. Miwa, H. Imada, M. Imai-Imada, K. Kimura, M. Galperin, and Y. Kim, *Many-Body State Description of Single-Molecule Electroluminescence Driven by a Scanning Tunneling Microscope*, Nano Lett. **19**, 2803 (2019).
[8]  H. Imada, K. Miwa, M. Imai-Imada, S. Kawahara, K. Kimura, and Y. Kim, *Real-Space Investigation of Energy Transfer in Heterogeneous Molecular Dimers*, Nature **538**, 364 (2016).
[9]  S. Jiang, T. Neuman, R. Bretel, A. Boeglin, F. Scheurer, E. Le Moal, and G. Schull, *Many-Body Description of STM-Induced Fluorescence of Charged Molecules*, Phys. Rev. Lett. **130**, 126202 (2023).
[10]  M. Imai-Imada et al., *Orbital-Resolved Visualization of Single-Molecule Photocurrent Channels*, Nature **603**, 829 (2022).
[11]  A. W. Frenkel, *MULTIPLICITY OF ELECTRON TRANSPORT REACTIONS IN BACTERIAL PHOTOSYNTHESIS*, Biol. Rev. **45**, 569 (1970).
[12]  J. Repp, G. Meyer, F. E. Olsson, and M. Persson, *Controlling the Charge State of Individual Gold Adatoms*, Science **305**, 493 (2004).
[13]  I. Swart, T. Sonnleitner, and J. Repp, *Charge State Control of Molecules Reveals Modification of the Tunneling Barrier with Intramolecular Contrast*, Nano Lett. **11**, 1580 (2011).
[14]  C. Wagner, M. F. B. Green, P. Leinen, T. Deilmann, P. Krüger, M. Rohlfing, R. Temirov, and F. S. Tautz, *Scanning Quantum Dot Microscopy*, Phys. Rev. Lett. **115**, 026101 (2015).
[15]  S. Fatayer, B. Schuler, W. Steurer, I. Scivetti, J. Repp, L. Gross, M. Persson, and G. Meyer, *Reorganization Energy upon Charging a Single Molecule on an Insulator Measured by Atomic Force Microscopy*, Nat. Nanotechnol. **13**, 376 (2018).
[16]  W. Steurer, S. Fatayer, L. Gross, and G. Meyer, *Probe-Based Measurement of Lateral Single-Electron Transfer between Individual Molecules*, Nat. Commun. **6**, 8353 (2015).
[17]  V. Rai, L. Gerhard, Q. Sun, C. Holzer, T. Repän, M. Krstić, L. Yang, M. Wegener, C. Rockstuhl, and W. Wulfhekel, *Boosting Light Emission from Single Hydrogen Phthalocyanine Molecules by Charging*, Nano Lett. **20**, 7600 (2020).
[18]  S. W. Wu, G. V. Nazin, X. Chen, X. H. Qiu, and W. Ho, *Control of Relative Tunneling Rates in Single Molecule Bipolar Electron Transport*, Phys. Rev. Lett. **93**, 236802 (2004).
[19]  J. Repp, G. Meyer, S. M. Stojković, A. Gourdon, and C. Joachim, *Molecules on Insulating Films: Scanning-Tunneling Microscopy Imaging of Individual Molecular Orbitals*, Phys. Rev. Lett. **94**, 026803 (2005).
[20]  T. Ardhuin, O. Guillermet, A. Gourdon, and S. Gauthier, *Measurement and Control of the Charge Occupation of Single Adsorbed Molecules Levels by STM and Nc-AFM*, J. Phys. Chem. C **123**, 26218 (2019).

Acknowledgments

We thank Rolf Allenspach, Guillaume Schull, Anna Rosławska, Song Jiang, Kirill Vasilev, Daniel Wegner, Florian Albrecht, Shadi Fatayer, Fabian Paschke, Armin Knoll, Shantanu Mishra, and Stefan Fölsch for discussions and comments. This work was supported by the ERC Synergy Grant MolDAM (no. 951519), the EU FET-OPEN project SPRING (no. 863098), and the H2020-MSCA-ITN ULTIMATE (no. 813036).

Author contributions

L. G. and K. K. designed the experiment. K. K., L.-A. L. and L. G. performed the experiments. All authors discussed the results and wrote the manuscript.

Competing interests

The authors declare no competing interests.

Data availability

All data needed to evaluate the conclusions in this paper are present in the paper and/or the Supporting Information.

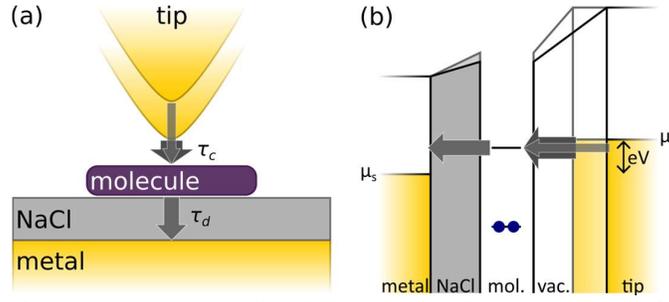

**Figure 1: Resonant tunneling through a molecule adsorbed on a thin NaCl film.** (a) Schematic depiction of the transient negative charging of the molecule at positive bias. The charging time through the vacuum barrier $\tau_c$ can be varied by changing the distance between tip and molecule; thicker arrows indicate smaller charging times. The discharging time, i.e., the charge-state lifetime, through the NaCl film $\tau_d$ is governed by the film thickness. The two cases that are schematically depicted here correspond to tip-molecule distances where $\tau_c$ is longer (long, thin arrow) and shorter (short, thick arrow) compared to $\tau_d$. (b) Corresponding one-electron picture of the double-barrier tunnel junction shown in (a).

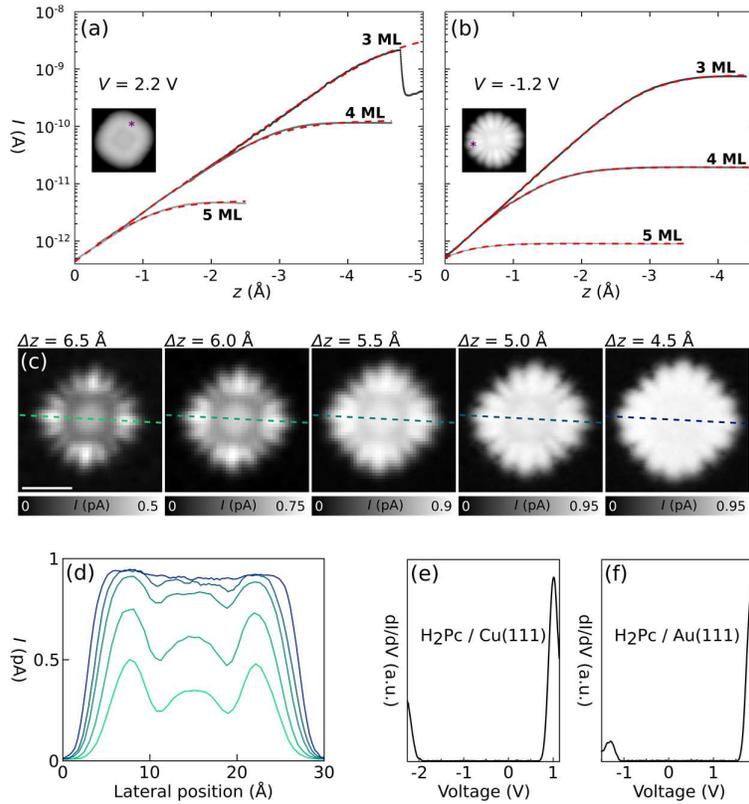

**Figure 2: Current as a function of tip height for ZnPc adsorbed on NaCl films of different thicknesses on Au(111).** $I(z)$ spectroscopy at the (a) negative ion resonance (NIR) and (b) positive ion resonance (PIR) of ZnPc molecules adsorbed on 3 - 5 monolayer (ML) NaCl/Au(111). Fits to the data are shown in red. On 3 ML at NIR, the molecule dislocated at $z \approx -4.8$ Å, resulting in an abrupt change in current. The insets in (a) and (b) exemplify the lateral position of the tip, atop regions of high orbital density, during the spectra. (c) Constant-height (c.h.) STM images ($V = -1.2$ V) of ZnPc adsorbed on 5 ML NaCl/Au(111) recorded at at different tip-sample distances. The tip-height offset $\Delta z$ is given with respect to the STM setpoint of $V = -1.2$ V, $I = 0.5$ pA above the bare NaCl surface. The scale bar corresponds to 1 nm and applies to all images in (c). (d) Line profiles along the dashed lines indicated in (c). (e, f) $dI/dV(V)$ of $H_2Pc$ adsorbed on 4 ML NaCl on Cu(111) (setpoint $V = 1.6$ V, $I = 0.5$ pA, $\Delta z = -1$ Å) (e) and 3 ML NaCl on Au(111) (setpoint $V = 2$ V, $I = 0.5$ pA, $\Delta z = -1$ Å) (f). The $dI/dV(V)$ was obtained by numerical differentiation of constant-height $I(V)$-curves.

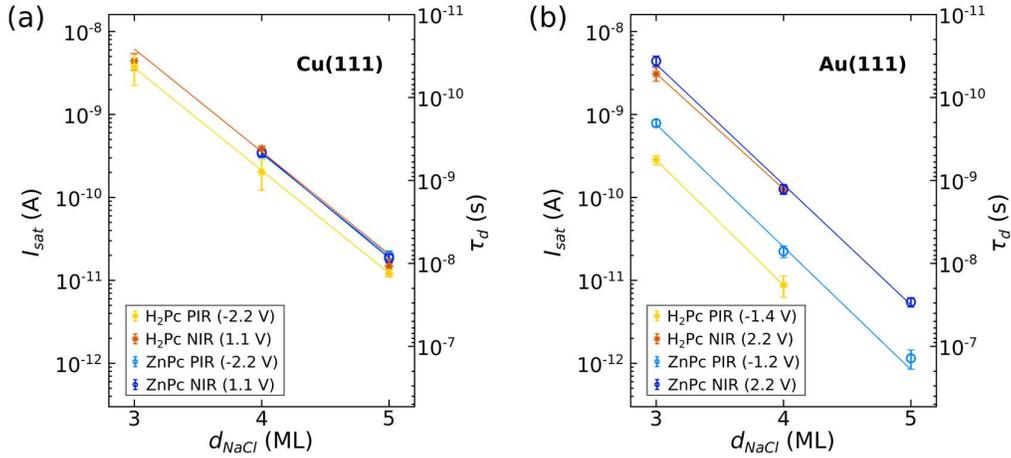

**Figure 3: Charge-state lifetimes.** Measured saturation currents $I_{sat}$ (left axis, log scale) and corresponding charge-state lifetimes $\tau_d$ (right axis, inverted log scale) of ZnPc and H$_2$Pc on NaCl on (a) Cu(111) and (b) Au(111) as a function of NaCl thickness. The saturation current was recorded at voltages within the ion resonances. The voltages that were typically used to record the $I(\Delta z)$-curves are indicated in the legends. Their absolute values are slightly larger than the voltages, at which the corresponding resonance is seen in d$I$/d$V$. The solid lines are fits to the data.

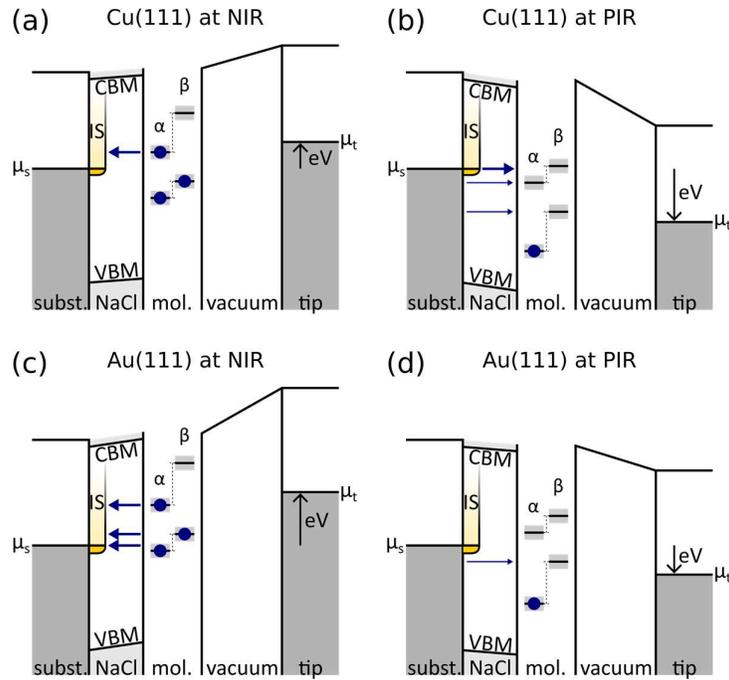

**Figure 4: Transition from a molecule's transiently charged state to the neutral charge state by tunneling through the NaCl barrier.** The frontier molecular levels, chemical potentials of tip ($\mu_t$) and sample ($\mu_s$), and the interface state (IS) are indicated. The shown level alignment corresponds to voltages of the ion resonances deduced from the positions of the neutral molecule's ion resonances and STML. The dashed lines indicate which single electron states derive from the HOMO (lower pair of states) and LUMO (higher pair of states) of the neutral molecule, the corresponding α and β spin channels are indicated. The grey shaded area indicates the linewidth of the levels. Because of different work functions, the vacuum-level aligned molecular ion resonances are shifted to larger bias values on Au(111) compared to Cu(111) [12,32]. Thicker (thinner) arrows indicate channels that involve (do not involve) the IS. In a)-c), the IS contributes to at least one channel, while in d), it does not. A more detailed depiction of how the shown level alignment was derived is shown in the SI in Fig. S5. By schematically showing the conduction band minimum (CBM) and valence band maximum (VBM) of NaCl we indicate the need to consider the NaCl's band structure for tunneling. References [66-68] suggest that the CBM is roughly aligned with the vacuum level while the band gap is at least 8 eV (see section S5 in the SI).

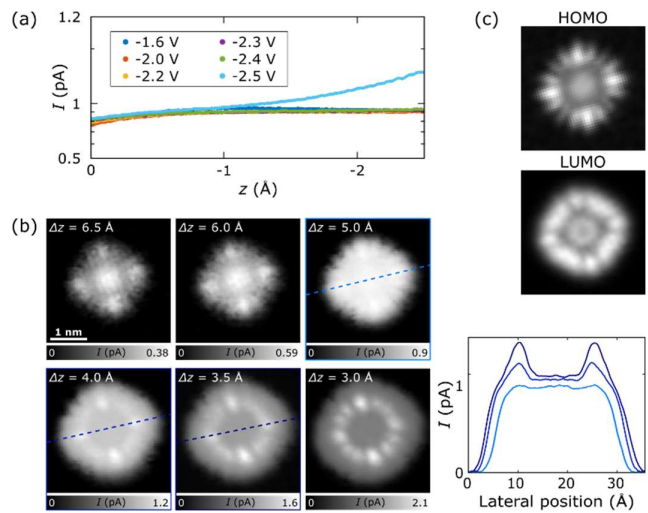

**Figure 5: Contribution of higher lying states of ZnPc on 5 ML NaCl/Au(111).** (a) $I(z)$-curves atop one of the lobes of the ZnPc HOMO at different $V$ from -1.6 V to -2.5 V. (b) C.h. STM maps on ZnPc adsorbed on 5 ML NaCl/Au(111) at different $\Delta z$ recorded at $V$ = -2.5 V. The dashed lines indicate the position of the line profiles shown in the right panel. The scale bar corresponds to 1 nm and applies to all images. $\Delta z$ is given with respect to the STM setpoint of $V$ = -2.5 V, $I$ = 0.5 pA above the bare NaCl surface. (c) C.h. STM images of ZnPc adsorbed on 5 ML NaCl/Au(111) at PIR (HOMO) and NIR (LUMO) for comparison.

# Supporting Information for

## Charge-state lifetimes of single molecules on ultrathin insulating films

Katharina Kaiser*[1] #, Leonard-Alexander Lieske[1], Jascha Repp[2], Leo Gross*[1]

[1] IBM Research Europe – Zurich, Säumerstrasse 4, 8803 Rüschlikon, Switzerland
[2] Department of Physics, University of Regensburg, Universitätsstraße 31, 93053 Regensburg, Germany
# Present address: Université de Strasbourg, CNRS, IPCMS, UMR 7504, F-67000 Strasbourg, France

*Corresponding authors. E-mail: katharina.kaiser@ipcms.unistra.fr, LGR@zurich.ibm.com

Contents:

S1. Additional constant-height current maps

S2. Distance and voltage dependence of the charge-state lifetime

S3. Decharging from the substrate

S4. Extraction of effective NaCl barrier heights

S5. Estimation of valence band maximum and conduction band minimum

S6. Supplementary references

## S1. Additional constant-height current maps

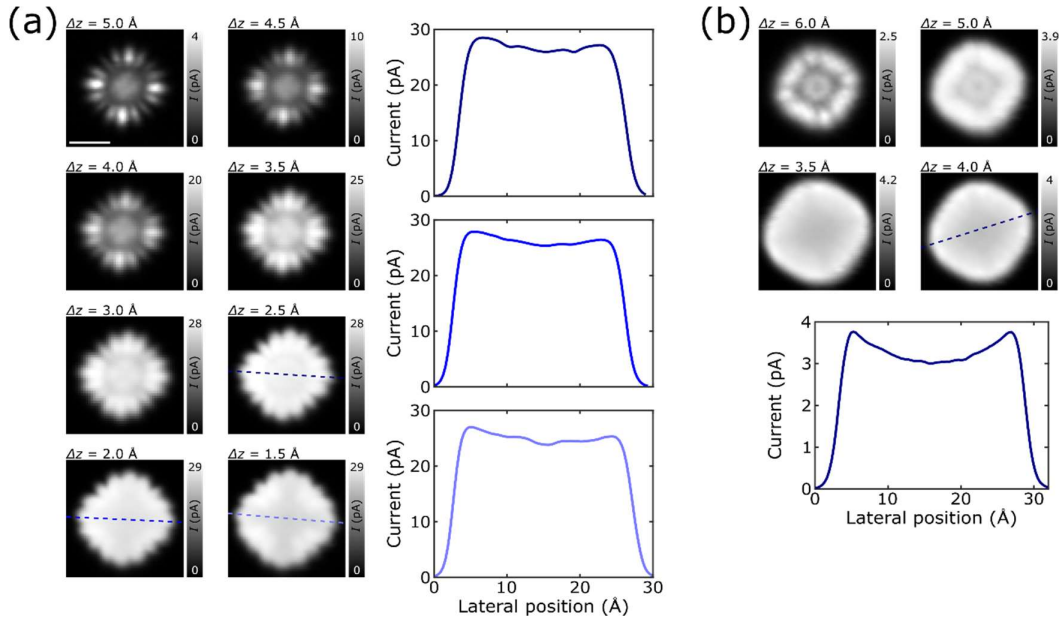

**Figure S1**: Constant height current maps at different tip-sample distances on ZnPc adsorbed on (a) 4 ML NaCl/Au(111), recorded at the PIR at $V$ = -1.2 V and (b) 5 ML NaCl/Au(111), recorded at the NIR at $V$ = 2.2 V. The line profiles correspond to current traces along the respective dashed lines at Δ$z$ = 2.5 Å, 2 Å and 1.5 Å in (a) as well as Δ$z$ = 4 Å in (b). Δ$z$ is given with respect to the STM setpoint of $V$ = -1.2 V, $I$ = 0.5 pA above the bare NaCl surface in (a) and $V$ = 2.2 V, $I$ = 0.5 pA in (b). The scale bar in (a) corresponds to 1 nm and is valid for all maps.

Figure S1 shows additional constant-height current maps recorded on ZnPc on 4 and 5 ML NaCl/Au(111). To ensure that the probed ionic resonance (PIR or NIR) is accessible at all probed tip heights and NaCl thicknesses, we used bias voltages $V$ slightly above (below) the peak position of NIR (PIR) in d$I$/d$V$. This is further detailed in section S2. In the constant-height STM images in the regime of current saturation (see linescans in Fig. S1), we observe a slightly larger saturation current at the rim of the molecule compared to its center. This effect probably results from a different screening of the tip's electric field by the molecule at different tip positions, slightly affecting the tunnel rates through the NaCl barrier.

## S2. Distance and voltage dependence of the charge-state lifetime

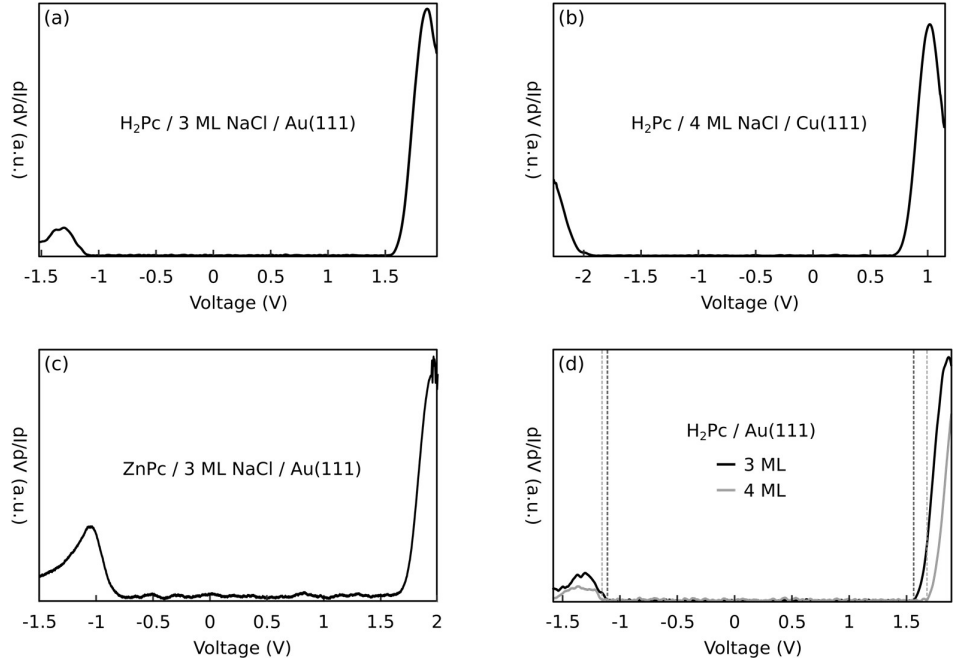

**Figure S2**: d$I$/d$V$($V$) recorded on different systems. (a) H$_2$Pc on 3 ML NaCl/Au(111) (setpoint $I$ = 0.5 pA, $V$ = 2 V, $\Delta z$ = -1 Å). (b) H$_2$Pc on 4 ML NaCl/Cu(111) (setpoint $I$ = 0.5 pA, $V$ = 1.6 V, $\Delta z$ = -1 Å). (c) ZnPc on 3 ML NaCl/Au(111). (d) Comparison of H$_2$Pc on 3 ML and on 4 ML NaCl/Au(111) (setpoint $I$ = 0.5 pA, $V$ = 2 V, $\Delta z$ = -1 Å). The onsets of PIR and NIR are indicated by the dashed lines.

The voltage drop across the tip-sample junction is composed of the voltage drop over the vacuum gap (between tip and molecule) and the one over the NaCl film (between molecule and metal substrate). The ratio of these two voltage drops depends on the exact geometry, *i.e.*, the tip-molecule distance $z$, the tip position, the thickness of the NaCl film $d_{NaCl}$, and the tip-radius. Following the estimation given in ref. [1], we estimate the relative voltage drop over NaCl to be in the range of 10-20% of the total applied bias voltage for the geometries present in this study.

Upon applying a bias, the voltages of the molecular ion resonances are shifted with respect to the zero bias condition by the voltage drop across the NaCl film. As a consequence, changes in $z$ or $d_{NaCl}$ directly affect the voltages of the ion resonances. To avoid shifting of the molecular ion resonances outside the voltage window when changing the lever arm ($z$ or $d_{NaCl}$), we typically used absolute bias voltages slightly larger (up to a few 100 mV larger) then the corresponding peak positions in d$I$/d$V$, $V_{PIR}$ and $V_{NIR}$. One example for the shift due to the changed lever arm is shown in Fig. S2d, where we show d$I$/d$V$($V$)-curves recorded atop H$_2$Pc adsorbed on 3 and 4 ML of NaCl on Au(111). The peaks and onsets of PIR and NIR shift to higher absolute values on 4 ML NaCl, as a result of the larger relative voltage drop across NaCl on 4 ML compared to 3 ML.

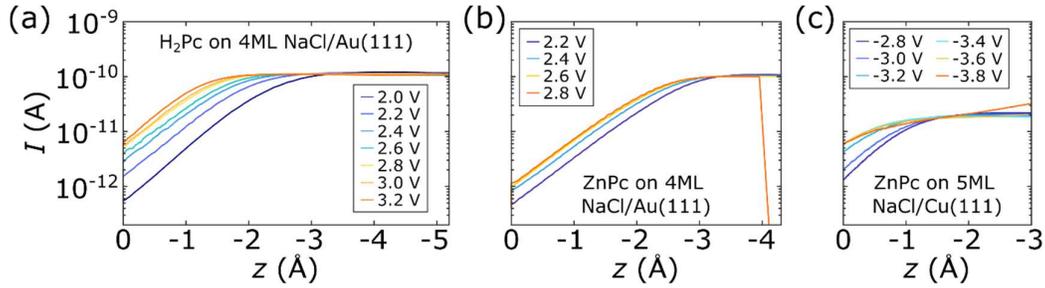

**Figure S3:** *I(z)* spectroscopy at different *V*. (a) *I(z)* at different positive bias voltages atop a lobe of an $H_2Pc$ and (b) ZnPc adsorbed on 4 ML NaCl/Au(111). In (b), the ZnPc molecule dislocated at *V* = 2.8 V at about *z* = -4 Å. (c) *I(z)* at different negative bias voltages atop one lobe of a ZnPc adsorbed on 5ML NaCl on Cu(111). Here, *z* corresponds to the tip-height offset Δ*z* when the STM feedback was opened, *i.e.*, at the position the spectrum was recorded at *I* = 0.5 pA, and *V* = 2.0 V (a), *V* = 2.2 V (b), and *V* = -2.8 V (c).

A similar trend is expected for decreasing *z* (*i.e.*, a decrease in *z* will lead to an increase in voltage drop across NaCl and thus an up-shift of the absolute values of $V_{PIR}$ and $V_{NIR}$) [2]. Unaffected by the changed lever arm, the saturation current, *i.e.*, the charge-state lifetime, does not vary with decreasing *z* (see, *e.g.*, Fig. 2b in the main text). In addition, we recorded bias dependent *I(z)* curves (Fig. S3), showing that, also independent of the applied bias voltage, the charge-state lifetime remains the same. This indicates that the charge-state lifetime, and with it the barrier height for the corresponding tunneling process through NaCl, does not vary significantly as a function of the voltage drop across NaCl.

However, for significantly increased voltages, see Fig. S3c at *V* = 3.8 V, the saturation current becomes affected by the increased voltage. This indicates that for ZnPc on Cu(111) for V ≤ -3.8 V additional tunnel channels become accessible. Figure S4 shows higher-lying states that could potentially contribute to the overall tunnel current at elevated bias voltages (*i.e.*, -3.8 V on Cu(111), Fig. S3c, and -2.5 V on Au(111), Fig. 5) in an energy diagram including many-body transitions. Note that here we do not discuss transitions to states that become directly accessible from $S_0$ in a single-electron tunneling event (faded arrows in Fig. S4) for the following reason: The increase in current above $I_{sat}$ and change in contrast in c.h. STM only appears at small *z*, *i.e.*, sets in only at a large tunnel rate between tip and molecule. This indicates that the underlying process is of higher order (*i.e.*, a two- or more-electron process) and involves tunneling into a transient state of the molecule (*e.g.*, $T_1$ or $D_0^+$) and hence only becomes accessible if charge transfer from the tip happens at comparable or faster time scales than the depopulation of the involved transient state.

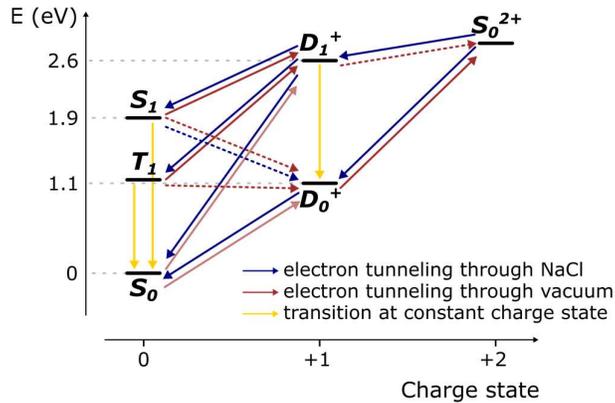

**Figure S4**: Energy diagram of ZnPc on Au(111), including higher-lying states and possible many-body transitions that become accessible at increased absolute voltage values. Blue arrows indicate charge-state transitions by tunneling between molecule and sample (through the NaCl barrier), and red arrows by tunneling between molecule and tip (through the vacuum barrier). Single-electron transitions from the neutral ground state $S_0$ (faded red lines) are not considered for the observed increase in current at small z. Yellow arrows indicate transitions that do not involve a change in charge state, such as radiative transitions. The dashed red arrows indicate charge-state transitions in which electrons are removed from higher-lying orbitals (the LUMO of the neutral molecule). These transitions could lead to the observed contrast in constant-height STM images at *V* = -2.5 V and small tip height. Note that this diagram is simplified and does not contain additional, even higher-lying excited states of the system.

At elevated bias voltages and for small $z$, charge transfer from tip to molecule could thus lead to transitions to $S_0^{2+}$, from which transitions to $D_1^+$ are possible, as indicated in Fig. S4, which can rationalize the observed increase in saturation current. Furthermore, the observation of the shape and symmetry of the LUMO density could relate to charge state transitions from the tip that involve higher-lying orbitals, as indicated by the red dashed lines in Fig. S4. The observed increased tunnel current in Fig. 5, which resembles the shape of the LUMO, could be caused, for example, by alternating charge transitions between the $D_1^+$ and $S_0^{2+}$ states.

**S3. Decharging from the substrate**

Figure 4 in the main text schematically depicts, in a single-electron picture, the molecular energy levels of ZnPc/H$_2$Pc after charging from the tip. Note that the single-electron picture is used here to depict the level alignment of the channels with respect to, *e.g.*, VB, CB and IS. Although the energy levels derive from the neutral molecule's HOMO and LUMO, their energetic positions are different from the neutral molecule's levels because of, *e.g.*, Coulomb interaction, the lifted spin degeneracy, and reorganization [1,3]. The exact energetic positions of these levels, corresponding to transiently charged states, cannot be probed in our experiment. However, we can estimate the energetic position of the cation's (anion's) singly unoccupied (occupied) frontier orbital with respect to the tip's electrochemical potential, based on the positions of the PIR (NIR) peak in d$I$/d$V$ and estimating the reorganization energy. From these levels (SOMO of anion at NIR and SUMO of cation at PIR), we construct the levels corresponding to the transitions to $S_1$ and $T_1$ states by using reported energies of luminescence and phosphorescence, respectively. The positions of the lowest unoccupied level at NIR and the highest occupied level at PIR are estimated from experiments in which molecules were doubly negatively charged on thick NaCl films [4], with consideration of the lever arm [3]. The charging energy $E_{charge2}$ for doubly charged states, corresponding to the additional energy needed from the single charging event to double charging, is estimated here as $E_{charge2}$ = 1.2 eV for both dianions and dications.

In addition, to corroborate the accessibility of possible channels to excited states in the neutralization step, we compared our conclusions to previous STML experiments on these systems. If a molecule shows luminescence in STML at PIR or NIR (on a given surface) the transition from the corresponding charged $D_0$ state back to the neutral charge state can entail the transition to $S_1$ [5–8]. Hence, if at a given bias voltage luminescence can be observed for the system, the channel to the $S_1$ state is accessible for the neutralization of the molecule.

Based on these arguments, we sketch in Fig. 4 the energetic positions of the molecular single-electron energy levels of the charged molecule with respect to tip and sample states, taking into account several literature values and observations from STML experiments, indicating possible pathways for the neutralization of the singly charged systems.

Figure S5 depicts the quantities that were taken into account for the level alignment presented in Fig. 4 in the main text, using the examples of the anion at bias voltages that correspond to the NIR (Fig. S5a) and the cation at voltages corresponding to the PIR (Fig. S5b) on Au(111). The sample voltages for these conditions, *i.e.*, at NIR and at PIR, correspond to the respective peaks in d$I$/d$V$. As a result of reorganization, the anion's (cation's) energy levels are shifted down (up) by the reorganization energy $E_{reorg}$ with respect to the corresponding energy levels of the neutral molecule. We assume here a reorganization energy of $E_{reorg}$ = 0.4 eV. The energy levels are broadened due to electron-phonon coupling [2,9]; the total peak width, being the energy range within which tunneling into the molecular orbitals is appreciable in the experiment, is here assumed to be $w$ = 0.6 eV. The energetic difference between the molecular levels are associated with the transition energies of the $S_0$-$S_1$ and $S_0$-$T_1$ transitions, respectively.

The energies for the $S_1 \rightarrow S_0$ transition in H$_2$Pc and ZnPc are 1.81 eV and 1.89 eV, respectively [5,10]. $T_1$ energies of 1.24 eV for H$_2$Pc and 1.14 eV for ZnPc have been reported for molecules in chloronaphthalene solution [11,12]. Note, however, that the presence of the substrate and the tip is known to shift the energies of optical transitions with respect to the values in solution [13,14], and phosphorescence from $T_1$ of H$_2$Pc or ZnPc in STML has not been reported as of now.

Note that for the PIR on Au(111), the applied bias voltages are close to being sufficient to facilitate the formation of the $T_1$ upon neutralization from the substrate. However, the exact energies of the $T_1 \rightarrow S_0$ transition for H$_2$Pc and ZnPc adsorbed on an ultrathin insulating film atop a metal substrate are not known. We tentatively excluded the tunnel channel for $T_1$ formation in Fig. 4d based on the energetic arguments given above. However, it cannot be fully ruled out because of the uncertainty of the energy of the $T_1 \rightarrow S_0$ transition and the significant level broadening on NaCl.

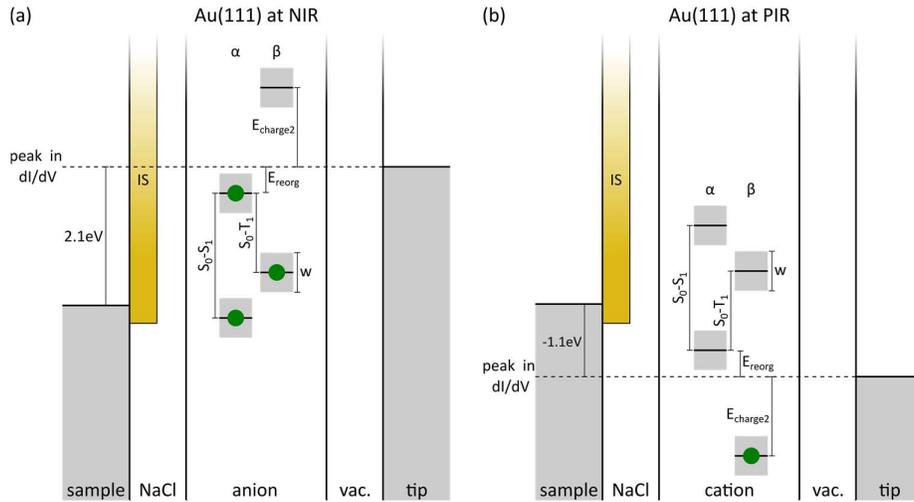

**Figure S5**: Sketch depicting the estimated quantitative alignment of the anionic (a) and cationic (b) energy levels with respect to sample and tip states. The assumed values for reorganization energy are $E_{reorg}$ = 0.4 eV, charging energy for doubly charged states $E_{charge2}$ = 1.2 eV, and linewidth of the energy levels $w$ = 0.6 eV, indicated by the grey shaded areas around the molecular levels. The energetic difference between the molecular levels are associated with the transition energies of the $S_0$-$S_1$ and $S_0$-$T_1$ transitions, respectively. The onset of the IS for NaCl/Au(111) is $V$ = -0.27 V. All shown energy differences are to scale.

The conduction band minimum (CBM) and valence band maximum (VBM) are indicated in Fig. 4 to highlight the need to consider the NaCl's band structure for tunneling. References [15–17] suggest that the CBM is roughly aligned with the vacuum level while the band gap is at least 8 eV. The NaCl film lowers the metal work function, *i.e.*, 4.9 eV for Cu(111) and 5.3 eV for Au(111), by about 1 eV [18,19]. Thus, the band gap ranges from around 4 – 5 eV below the electrochemical potential of the sample to around 3.5 – 4.5 eV above.

## S4. Extraction of effective NaCl barrier heights

For a single-barrier tunnel junction, the tunnel current as a function of distance $z$ is given by

$$I(z) = I_0 \cdot e^{-2\kappa z}$$

The exponent is determined by $\kappa = \frac{\sqrt{2m_e \Phi_{eff}}}{\hbar}$, with the electron mass $m_e$ and the effective barrier height $\Phi_{eff}$. For tunneling between the molecule and metallic substrate, the effective barrier height for tunneling through the NaCl film $\Phi_{NaCl}$ can be determined from the exponent of the $I_{sat}(d_{NaCl})$ dependence, *i.e.*, the slope $m$ of $\ln(I_{sat}(d_{NaCl}))$.

We determined slopes $m$ and effective NaCl barrier heights $\Phi_{NaCl}$ at both ion resonances for ZnPc and H$_2$Pc adsorbed on NaCl on Au(111) and Cu(111). The results are summarized in the following Table S1. Note that, for H$_2$Pc on Au(111) and ZnPc on Cu(111), the saturation current could only be recorded on two different NaCl film thicknesses and hence, the values determined for these systems have a larger uncertainty.

|  | $m$ at PIR | $m$ at NIR | Increase of $\tau_d$ per ML NaCl at PIR | Increase of $\tau_d$ per ML NaCl at NIR | $\Phi_{NaCl}$ at PIR | $\Phi_{NaCl}$ at NIR |
|---|---|---|---|---|---|---|
| ZnPc/Au(111) | -3.4 ± 0.2 | -3.3 ± 0.2 | x 30 ± 6 | x 28 ± 4 | 1.3 ± 0.2 eV | 1.3 ± 0.1 eV |
| H$_2$Pc/Au(111) | -3.5 ± 0.4 | -3.2 ± 0.3 | x 32 ± 11 | x 25 ± 6 | 1.4 ± 0.3 eV | 1.2 ± 0.2 eV |
| ZnPc/Cu(111) | -2.8 ± 0.2 | -2.9 ± 0.2 | x 17 ± 4 | x 19 ± 4 | 0.9 ± 0.2 eV | 1.0 ± 0.2 eV |
| H$_2$Pc/Cu(111) | -2.9 ± 0.1 | -2.9 ± 0.4 | x 17 ± 2 | x 17 ± 6 | 0.9 ± 0.1eV | 0.9 ± 0.2 eV |

**Table S1**: Summary of slopes $m$, increase in charge-state lifetime per additional ML NaCl, and corresponding effective NaCl barrier height for ZnPc and H$_2$Pc adsorbed on Cu(111) and Au(111) and recorded at PIR and NIR.

The saturation current was determined with the fitting procedure discussed in the manuscript. The corresponding uncertainty is estimated based on the standard deviation of the data points.

The uncertainty margin of the slopes was determined from the linear fit in the case of ZnPc/Au(111) and H$_2$Pc/Cu(111). For H$_2$Pc/Au(111) and ZnPc/Cu(111), where the saturation current was only measured for two different NaCl film-thicknesses, this uncertainty was determined from the standard deviation of the measured saturation current.

## S5. Estimation of valence band maximum and conduction band minimum

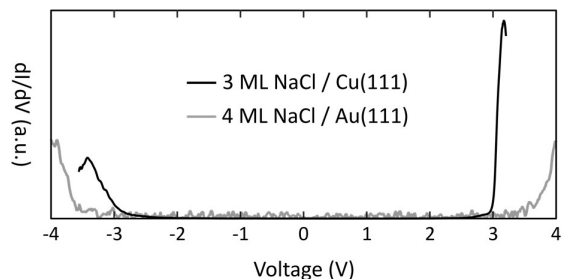

**Figure S6**: *dI/dV*(*V*) on 3 ML NaCl/Cu(111) (black curve, setpoint *V* = -0.2 V, *I* = 2 pA, *Δz* = 2 Å) and 4 ML NaCl/Au(111) (grey curve, setpoint *V* = -5 V, *I* = 2 pA, *Δz* = 0 Å). The curves were obtained by numerical differentiation of *I*(*V*) curves.

Figure S6 shows d*I*/d*V*(*V*) spectra recorded atop the bare NaCl surface for different underlying metal substrates. Note that the measured apparent onsets of tunnelling depend on the tip-NaCl distance and might relate to image potential states [20,21] and only serve as a rough lower bound for the band onsets. The combination of refs. [15–17] indicate that the CBM is located close to the vacuum level and that even for films of only few atomic layers the band gap of NaCl is (almost) fully developed, *i.e.*, about 8.5 eV. The NaCl film lowers the metal work function, *i.e.*, 4.9 eV for Cu(111) and 5.3 eV for Au(111), by about 1 eV [18,19]. We estimate that the band gaps, that is VBM and CBM, should extend from roughly 4 - 5 eV below the Fermi level to roughly 3.5 - 4.5 eV above it.